# Graphene field-effect-transistors with high on/off current ratio and large transport bandgap at room temperature

Fengnian Xia[*], Damon B. Farmer, Yu-ming Lin, and Phaedon Avouris[*]

*IBM Thomas J. Watson Research Center, Yorktown Heights, NY 10598*

Abstract. Graphene is considered to be a promising candidate for future nano-electronics due to its exceptional electronic properties. Unfortunately, the graphene field-effect-transistors (FETs) cannot be turned off effectively due to the absence of a bandgap, leading to an on/off current ratio typically around 5 in top-gated graphene FETs. On the other hand, theoretical investigations and optical measurements suggest that a bandgap up to a few hundred meV can be created by the perpendicular E-field in bi-layer graphenes. Although previous carrier transport measurements in bi-layer graphene transistors did indicate a gate-induced insulating state at temperature below 1 Kelvin, the electrical (or transport) bandgap was estimated to be a few meV, and the room temperature on/off current ratio in bi-layer graphene FETs remains similar to those in single-layer graphene FETs. Here, for the first time, we report an on/off current ratio of around 100 and 2000 at room temperature and 20 K, respectively in our dual-gate bi-layer graphene FETs. We also measured an electrical bandgap of >130 and 80 meV at average electric displacements of 2.2 and 1.3 V-nm$^{-1}$, respectively. This demonstration reveals the great potential of bi-layer graphene in applications such as digital electronics, pseudospintronics, terahertz technology, and infrared nanophotonics.

*To whom correspondence should be addressed.
Email: fxia@us.ibm.com (F.X.); avouris@us.ibm.com (P.A.).

Recently, graphene attracts enormous attention due to its unique electronic properties[1-3]. Creating a bandgap in graphene is probably one of the most important and tantalizing research topics in graphene community since it may ultimately enable new applications in digital electronics, pseudospintronics[4], terahertz technology[5], and infrared nanophotonics[6-9]. A number of approaches have been proposed or implemented to create a bandgap in single or bi-layer graphenes already, such as using uniaxial strain[10-11], graphene-substrate interaction[12], lateral confinement[13-17], and breaking the inversion symmetry in bi-layer graphenes[6, 18-23]. Graphene nano-ribbon field-effect transistors exhibit an electrical bandgap up to a few hundred meV and very large current on/off ratio even at room-temperature[15-17]. However, currently there is no reliable method to produce nano-ribbons with desirable nanometer scale width. Moreover, carrier mobility in ultra-narrow graphene nano-ribbons is usually not as high as that in large area graphene[16].

On the other hand, theoretical investigations predict a sizeable bandgap opening up to 300 meV in Bernal-stacking bi-layer graphene using a perpendicular E-filed to render the A1 and B2 sites (see Fig. 1a) nonequivalent[18-20]. Optical measurements did confirm a bandgap in bi-layer graphenes with broken inversion symmetry[6, 21, 23]. However, previous carrier transport measurements were not able to find such a large bandgap. Therefore, whether a considerable electrical (or transport) bandgap exists in biased bi-layer graphene remains an unsolved problem in graphene research. In our work, we recognized the importance of preserving the intrinsic properties of bi-layer graphene and introduced two important steps in the bi-layer graphene FET fabrication, which allowed us to observe a large electrical bandgap (> 130 meV) in biased bi-layer graphene.

Fig. 1b depicts a 3-dimensional schematic view of our dual-gate bi-layer graphene transistor and the layer structure within the channel region of such a device is shown in Fig. 1c. The bi-layer graphene channel is sandwiched completely between top and bottom gates. The bottom gate $SiO_2$ film is 300 nm thick. The top gate dielectric stack consists of first a 9±3 nm of an organic seed layer made from a derivative of polyhydroxystyrene (the polymer NFC 1400-3CP manufactured by JSR Micro, Inc.) followed by a 10±1 nm film of $HfO_2$ deposited by atomic layer deposition (ALD). Details of the fabrication processes are presented in the "Fabrication of the dual-gate bi-layer graphene FETs" section at the end of the main text and Ref. 24. This approach to FET fabrication allows us to probe the intrinsic properties of the biased bi-layer graphene and to observe a large electrical bandgap. The introduction of the organic seed layer before $HfO_2$ ALD not only facilitates the high-k gate $HfO_2$ deposition through methyl and hydroxyl groups contained within the polymer on the otherwise inert graphene surface, but also preserves the high mobility and intrinsic properties of the active graphene layer by reducing remote phonon and Coulomb scattering[24].

We first investigated the switching behavior our dual-gated bi-layer graphene FETs at room temperature. Figure 2a shows the transfer characteristics of a bi-layer graphene FET with a channel 1.6 µm wide by 3 µm long. In each curve, the back gate bias ($V_{bg}$) is fixed and the top gate bias ($V_{tg}$) is scanned from -2.6 to 6.4 V. $V_{bg}$ is varied from -120 to 80 V at steps of 20 V as shown by the black dashed arrow in Fig. 2a. The source is grounded and a drain bias of 1 mV is applied to the device. Here, both "on" and "off" currents are defined at a specific back gate bias. At different back gate biases, both on and off currents are different. However, we defined the device on/off current ratio to be the maximum current modulation factor possible at the optimum back gate bias (-120 V in this device) when modulating the top gate bias. A minimum off current of around 10 nA is realized at $V_{bg}$ and $V_{tg}$ of -120 and 6.4 V, respectively, as shown by the black curve in Fig. 2a, corresponding to a device on/off current ratio of about 100 at room temperature. In comparison, an on/off current ratio of about 4 is observed in a single layer graphene FET with similar device structure as shown in Fig. 2b. Hence the on/off current ratio in our bi-layer graphene FET is enhanced by a factor of 25 when compared with that of singly layer graphene FET. Moreover, the decrease of the off-current in our bi-layer graphene FET does not seem to cease at $V_{bg}$ and $V_{tg}$ of -120 and 6.4 V, respectively as shown by Fig. 2a. Further enhancing the top and back gate biases would result in even smaller off-current. However, in our current devices, this was not possible due to the limited strength of our gate dielectric stacks.

Scanning of the top gate bias not only modulates the doping of the bi-layer graphene, but also changes the induced bandgap[6, 21]. In each curve, at the minimum conductance the graphene sheet is approximately at the charge-neutrality condition. If an appreciable bandgap exists at this condition, the off-current would be dominated by the thermionic emission of carriers through the metal-graphene Schottky barrier. Hence the off-current, $I_{off}$, would be proportional to $\exp(-\frac{q\phi_{barrier}}{kT})$, where q is the electron charge, $\phi_{barrier}$ is the Schottky barrier height, k is the Boltzman constant, and T is the temperature. A maximum Schottky barrier height, $\phi^0_{barrier}$, is attained at the charge-neutrality point when the top and back gate biases are -120 and 6.4 V, respectively. The variation of the Schottky barrier, $\Delta(\phi_{barrier}) = \phi_{barrier}^0 - \phi_{barrier}$, can be inferred using the off-current at each charge-neutrality condition in each curve. The results are plotted in the inset of Fig. 2a as a function of the average electrical displacement, $D_{ave}$[6]. At the charge-neutrality condition, $D_{ave} \approx \varepsilon_{SiO_2}(V_{bg} - V_{bg0})/d_{SiO_2}$, where $\varepsilon_{SiO_2}$ (~ 3.9) is the dielectric constant of the back gate oxide, $V_{bg0}$ is the Dirac offset voltage (50 V in this device), and $d_{SiO_2}$ (300 nm) is the thickness of the back gate oxide. This approach is reliable when the Schottky barrier height is much larger than kT. When the barrier height is smaller or comparable with kT, the off-current is no longer limited by thermionic emission but also significantly affected by the metal-graphene contact and graphene channel resistances, leading to a reduced off-current. Therefore, the inset of Fig. 2a represents a lower limit to the Schottky barrier height created in the bi-layer graphene device. We can therefore conclude that in this bi-layer graphene, at an average electrical displacement of 2.2 Vnm$^{-1}$, the electrical bandgap is > 130 meV, assuming the Schottky barrier height is about half of the electrical bandgap. At a similar bias condition, the optical measurements indicated an optical bandgap of around 200 meV [6, 21].

The room temperature output characteristics of the same bi-layer graphene device in Fig. 2a are shown in Fig. 2c. In this measurement, the back gate bias ($V_{bg}$) is fixed at -100 V. The top gate bias ($V_{tg}$) is varied from -2 to 6 V at steps of 1 V. The inset of Fig. 2c depicts an enlarged view of the output characteristics at $V_{bg}$ = -100 V and $V_{tg}$ = 6 V. Current saturation is observed only in this curve, which shows a typical rectifying behavior of a metal-semiconductor junction[25] and implies again an appreciable bandgap opening in graphene at this biasing condition. By contrast, other curves in Fig. 2c show a completely linear behavior, which is typical for metal-zero-gap-semiconductor (graphene) junctions as reported in many previous publications.

We have also performed transfer characteristic measurements on graphene FETs as a function of temperature. Fig. 3a shows the transfer characteristics of another of these devices with a channel geometry of 1.2 µm wide by 1.5 µm long at 20 K. $V_{bg}$ is varied from 0 to 120 V at steps of 20 V as shown by the black dashed arrow in the inset of Fig. 3a. A device on/off current ratio of about 2000 is achieved at a fixed back gate bias of 120 V as shown by the black curve in Fig. 3a. At the minimum conductance ($V_{bg}$ = 120 V and $V_{tg}$ = -3. 5 V), the $D_{ave}$ is 1.3 Vnm$^{-1}$ since $V_{bg0}$ in this device is about 20 V. Fig. 3b shows the device on/off current ratio measured at 10 different temperatures with off-currents all taken at $D_{ave}$ of around 1.3 Vnm$^{-1}$. The improvement in device on/off current ratio at low temperature is due to the reduction in off-current. The off-current, $I_{off}$, for thermionic injection is proportional to $\exp(-\frac{q\phi_{barrier}}{kT})$ as discussed above, thus we may expect that $\ln(\frac{I_{on}}{I_{off}})$ vs. $\frac{1}{T}$ would yield a straight line with a slope of $\frac{q\phi_{barrier}}{k}$ [13]. In fact, the dashed line in Fig. 3b obtained from the on/off ratios from 295 to 100 K leads to a Schottky barrier height of 40 meV, corresponding to a electrical bandgap of 80 meV, if we assume the barrier height is about half of the gap. This barrier is smaller than the optical gap of ~130meV measured with optical techniques at a similar bias[6, 21]. Moreover, we observe that at temperatures below about 100 K, the on/off current ratio does not improve as fast as indicated by the dashed line, most likely due to the presence of tunnelling through defect states, i.e. tails in the density of states[13, 26]. This phenomenon has been observed in both carbon nanotube[26] and graphene nanoribbon[13] transistors. The optical measurements[6, 21] on the other hand reflect the peak in the joint density of states. Therefore, differences in electrical and optical bandgaps should not be surprising.

Finally, we note that in our current devices, the large conducting plate (back-gate) underneath the bi-layer graphene will decrease the operational speed of the device. For realistic applications, local bottom gates should be introduced to minimize such parasitic capacitances. Moreover, the on-current of the graphene transistors can be further enhanced by reducing the graphene-metal contact resistance and optimizing the device geometry. The off-current can be further suppressed by improving the overall gate dielectric strength (or overall gate dielectric constant) and the purity of bi-layer graphene. Hence, room temperature on/off current ratio of 100 is by no means the upper limit of the graphene FET.

In summary, we demonstrated a bi-layer graphene transistor with an on/off current ratio of around 100 at room temperature. The transport measurement indicates a Schottky barrier height >65 meV at $D_{ave}$ of 2.2 Vnm$^{-1}$, corresponding to an electrical (transport) bandgap of >130 meV. At 20 K, a device on/off current ratio of about 2000 is demonstrated at $D_{ave}$ of 1.3 Vnm$^{-1}$. Revealing of the large electrical bandgap in bi-layer graphene may enable a number of novel nanoelectronic and nanophotonic applications.

**Fabrication of the dual-gate bi-layer graphene FETs**

The fabrication steps of the dual-gate bi-layer graphene field effect transistor (FET) are described as follows:

1. Identification of bi-layer graphene flakes using optical approach and Raman spectroscopy. The bi-layer graphene flakes in this experiment were purchased from Graphene Industries, Inc.

2. First e-beam lithography and source/drain metallization (Ti/Pd/Au/Ti: 0.5/20/20/5 nm).

3. Second e-beam lithography and patterning of the bi-layer graphene channel.

4. Spin coating of the organic seed layer made from a derivative of polyhydroxystyrene (the polymer NFC 1400-3CP manufactured by JSR Micro, Inc.) for atomic layer deposition (ALD). The layer thickness can be adjusted by spin speed. The dielectric constant of this material is about $2.5^{24}$.

5. Atomic layer deposition of top gate oxide ($HfO_2$) at $T < 200^0C$.

6. Third e-beam lithography and top gate metallization (Ti/Au: 5/25 nm).

Poly methyl methacrylate (PMMA) was used as the e-beam resist in all the processing steps mentioned above. Removal of PMMA was realized using acetone and usually was followed by isopropanol rinse. No specific surface cleaning steps were involved in the processing.

**References**

1. Novoselov, K. S.; Geim, A. K.; Morozov, S. V.; Jiang, D.; Zhang, Y.; Dubonos, S. V.; Grigorieva, I. V.; Firsov A. A. *Science* **2004**, *306*, 666.

2. Novoselov, K. S.; Geim, A. K.; Morozov, S. V.; Jiang, D.; Katsnelson, M. I.; Grigorieva, I. V.; Dubonos, S. V.; Firsov, A. A. *Nature* **2005**, *438*, 197.

3. Zhang, Y.; Tan, J. W.; Stormer, H. L.; Kim, P. *Nature* **2005**, *438*, 201.

4. San-Jose, P.; Prada, E.; McCann, E.; Schomerus, H. *Phys. Rev. Lett.* **2009**, *102* 247204.

5. Tonouchi, M. *Nature Photon.* **2009**, *1*, 97.

6. Zhang, Y.; Tang, T.; Girit, C.; Hao, Z.; Martin, M. C.; Zettl, A.; Crommie, M. F.; Shen, Y. R.; Wang, F. *Nature* **2009**, *459*, 820.

7. Ryzhii, V.; Mitin, V.; Ryzhii, M.; Ryabova, N.; Otsuji, T. *Appl. Phys. Express* **2008**, *1*, 063002.

8. Wang, F.; Zhang, Y.; Tian, H.; Girit, C.; Zettl, A.; Crommie, M.; Shen, Y. R. *Science* **2008**, *320*, 206.

9. Xia, F.; Mueller, T.; Lin, Y-M.; Valdes-Garcia, A.; Avouris, Ph. Published online on 10/11/2009 in *Nature Nano.*, DOI: 10.1038/NNANO.2009.292.

10. Ni, Z. H.; Yu, T.; Lu, Y. H.; Wang, Y. Y.; Feng, Y. P.; Shen, Z. X. *ACS Nano* **2008**, *2*, 2301.

11. Pereira, V. M.; Castro Neto, A. H.; Peres, N. M. R. *Phys. Rev. B* **2009**, *80*, 045401.

12. Zhou, S. Y.; Gweon, G.–H.; Fedorov, A. V.; First, P. N.; de Heer, W. A.; Lee, D. –H.; Guinea, F.; Castro Neto, A. H.; Lanzara, A. *Nature Mat.* **2007**, *6*, 770.

13. Chen, Z.; Lin, Y. M.; Rooks, M. J.; Avouris, Ph. *Physica E* **2007**, *40*, 228.

14. Han, M. Y.; Ozyilmaz, B.; Zhang, Y.; Kim, P. *Phys. Rev. Lett.* **2007**, *98*, 206805.

15. Li, X.; Wang, X.; Zhang, L.; Lee, S.; Dai, H. *Science* **2008**, *319*, 1229.

16. Wang, X.; Ouyang, Y.; Li, X.; Wang, H.; Guo, J.; Dai, H. *Phys. Rev. Lett.* **2008**, *100*, 206803.

17. Ritter, K. A.; Lyding, J. W. *Nature Mat.* **2009**, *8*, 235-242.

18. McCann, E. *Phys. Rev. B* **2006**, *74*, 161403(R).

19. Castro, E. V.; Novoselov, K. S.; Morozov, S. V.; Peres, N. M. R.; Lopes dos Santos, J. M. B.; Nilsson, J.; Guinea, F.; Geim, A. K.; Castro Neto, A. H. *Phys. Rev. Lett.* **2007**, *99*, 216802.

20. Min, H.; Sahu, B.; Banerjee, S. K.; MacDonald, A. H. *Phys. Rev. B* **2007**, *75*, 155115.

21. Mak, K. F.; Lui, C. H.; Shan, J.; Heinz, T. F. *Phys. Rev. Lett.* **2009**, *102*, 256405.

22. Oostinga, J. B.; Heersche, H. B.; Liu, X.; Morpurgo, A. F.; Vandersypen, L. M. K. *Nature Mater.* **2008**, *7*, 151.

23. Ohta, T.; Bostwick, A.; Seyller, T.; Horn, K.; Rotenberg, E. *Science* **2006**, *313*, 951.

24. Farmer, D. B.; Chiu, H-Y.; Lin, Y-M.; Jenkins, K. A.; Xia, F.; Avouris, Ph. Published online on 11/02/2009 in *Nano Lett.*, DOI: 10.1021/nl902788u.

25. Sze, S. M. *Physics of semiconductor devices*, Ch.5 (Wiley, 1993).

26. Appenzeller, J.; Radosavljevic, M.; Knoch, J.; Avouris, Ph. *Phys. Rev. Lett.* **2004**, *92*, 048301.

**Acknowledgments**

The authors are grateful to B. Ek and J. Bucchignano for help in technical assistance. F. X. is indebted to C. Y. Sung for his encouragement. Correspondence should be addressed to F.X. (fxia@us.ibm.com) or P.A. (avouris@us.ibm.com).

**Figure Captions**

**Figure 1 Structure of the bi-layer graphene FET**

a.  A schematic view of bi-layer graphene in Bernal stacking. A1 and B2 are equivalent without vertical E filed shown by the green arrow hence the system posses inversion symmetry. This symmetry is broken under E-filed.
b.  A 3-dimensional schematic view of the dual-gate bi-layer graphene FET.
c.  The layer structure within this bi-layer graphene FET channel. From top to bottom are: top metal gate, $HfO_2$ deposited using ALD, organic polymer NFC 1400-3CP as ALD seeding layer, bi-layer graphene, bottom gate oxide, and silicon back gate. Red and blue dots denote carbon atoms. The thickness of the layers is not drawn to scale.

**Figure 2 Transport characteristics of a bi-layer graphene FET at room temperature**

a.  The room temperature transfer characteristics of a dual-gate bi-layer graphene FET. $V_{bg}$ is varied from -120 to 80 V at steps of 20 V. Inset: variation of the Schottky barrier height, $\Delta(\phi_{barrier})$, as a function of the average electrical displacement, $D_{ave}$, inferred from the off-currents at the charge-neutrality point.
b.  The room temperature transfer characteristics of a similar dual-gate, single layer graphene FET for comparison purpose. $V_{bg}$ is varied from -90 to 90 V at steps of 20 V. In this device, an on/off current ratio of 4 is obtained, 25-fold smaller if compared with that in dual-gate bi-layer graphene FET as shown in Fig. 2a.
c.  The room temperature output characteristics of the bi-layer graphene FET in Fig. 2a at $V_{bg}$ = -100 V and $V_{tg}$ from -2 to 6 V. Inset: enlarged view of the output characteristics at $V_{bg}$ = -100 V and $V_{tg}$ = 6 V. The horizontal and vertical axes are identical to those in the main figure.

**Figure 3 Transfer characteristics of a bi-layer graphene FET at 10 different temperatures**

a. The transfer characteristics of another bi-layer graphene FET at 20 K.
b. The device on/off current ratio vs. temperature at $D_{ave}$ of 1.3 Vnm$^{-1}$. The dashed line denotes the estimated on/off ratio vs. temperature for a Schottky barrier height of 40 meV, which corresponds to an electrical bandgap of 80 meV.

**Figure 1**

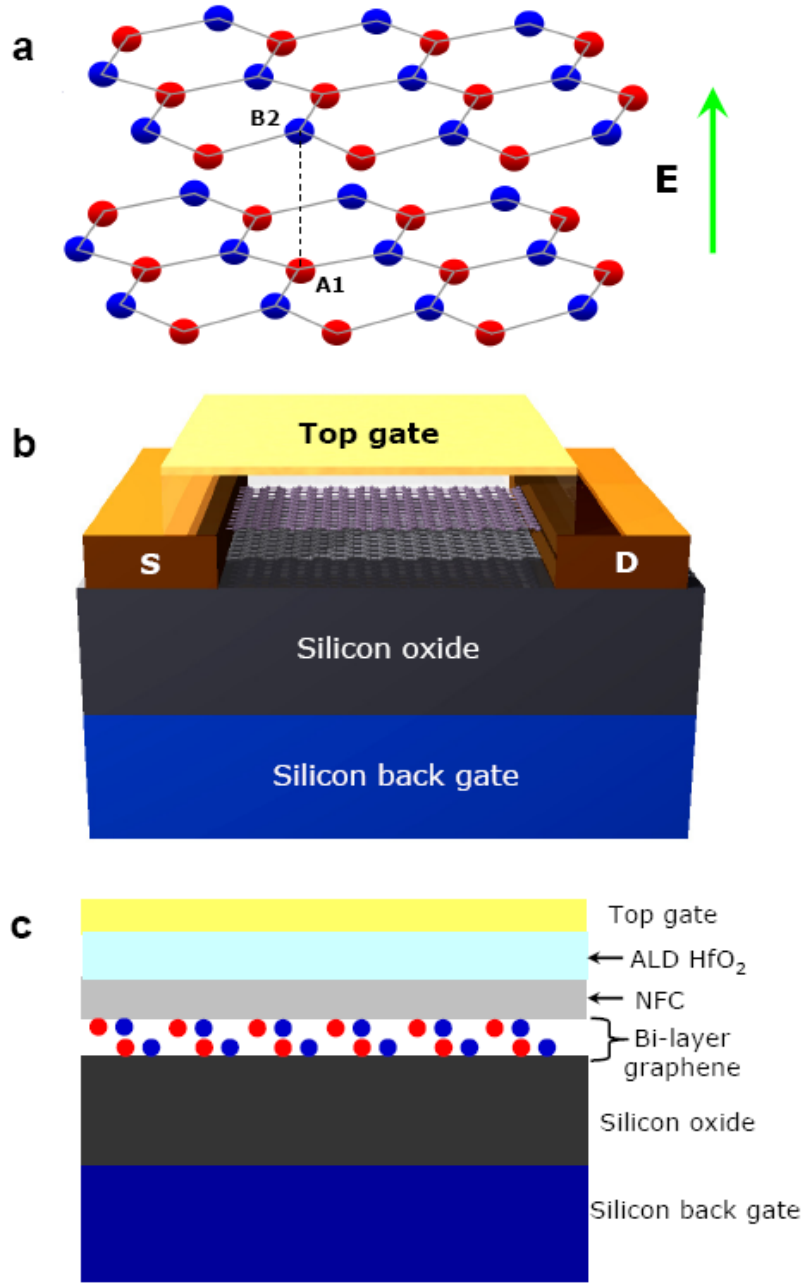

**Figure 2**

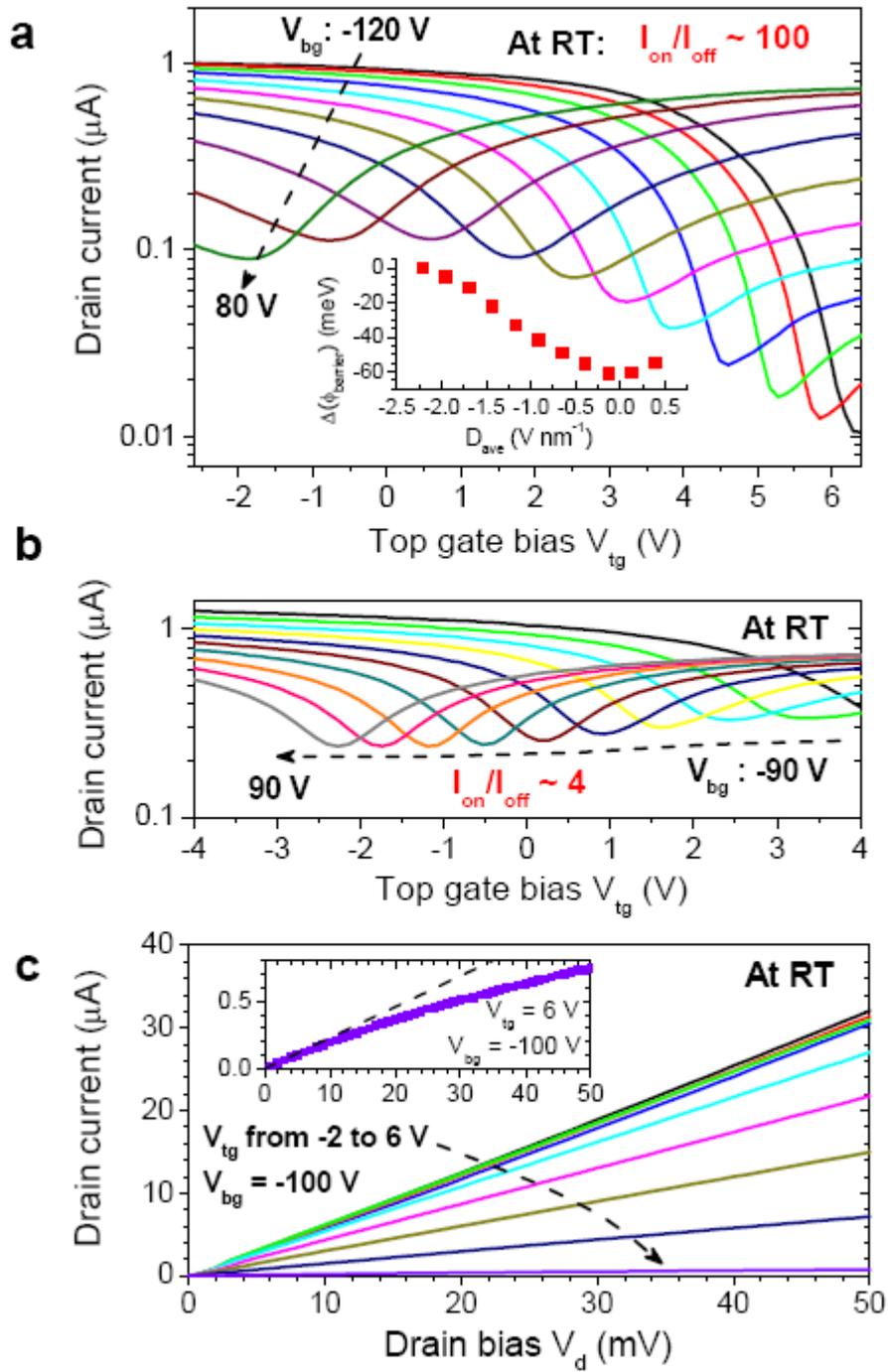

**Figure 3**

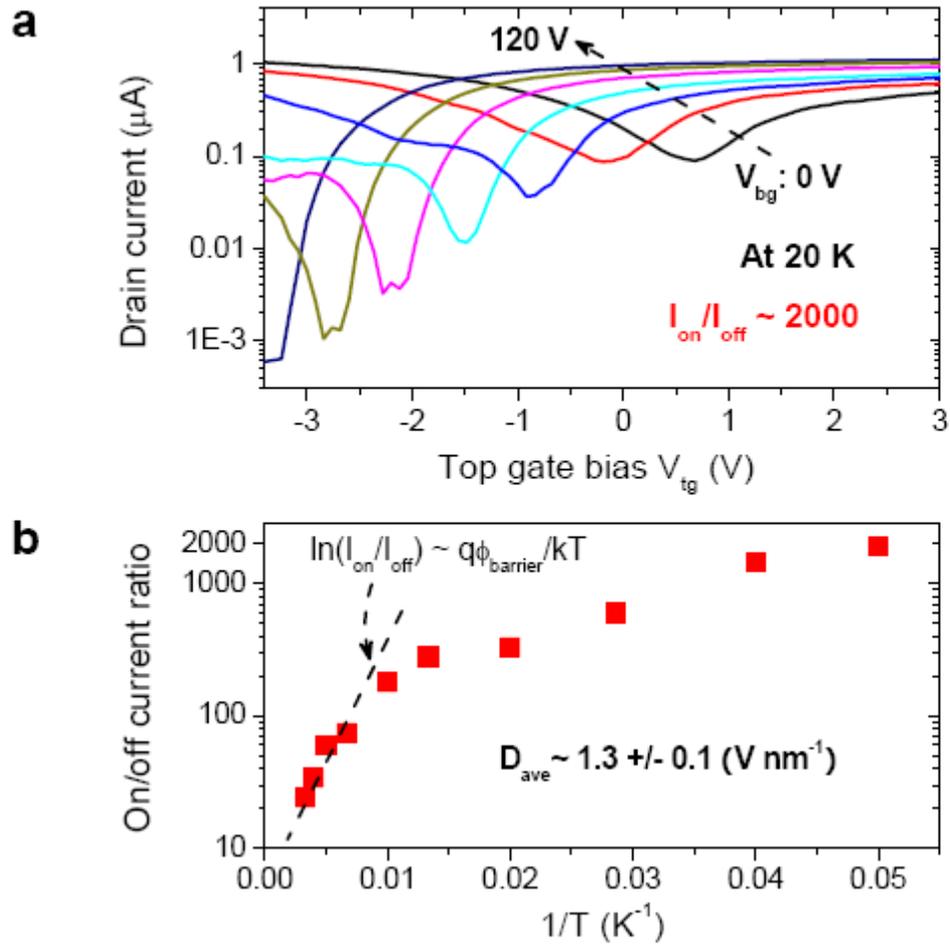